# Automated Software Tool for Compressing Optical Images with Required Output Quality

Sergey Krivenko, Alexander Zemliachenko, Vladimir Lukin, Alexander Zelensky

Dept of Transmitters, Receivers and Signal Processing, National Aerospace University named after N.Ye. Zhukovsky, UKRAINE, Kharkov, Chkalova Street, 17, E-mails: lukin@ai.kharkov.com, azelen@mail.ru, krivenkos@inbox.ru

*Abstract* – **the paper presents an automated software tool for lossy compression of grayscale images. Its structure and facilities are described. The tool allows compressing images by different coders according to a chosen metric from an available set of quality metrics with providing a preset metric value. Examples of the tool application to several practical situations are represented.**

*Keywords* – **image lossy compression, required quality, software implementation**

## I. Introduction

Compression of optical images including graphical ones used in CAD systems is one of the most typical and important tasks in many applications of digital image processing. As it is known, lossless compression rarely provides compression ratio (CR) larger than 1.5…2, and this is mostly inappropriate for typical applications. Due to this, lossy compression has become a standard tool in digital photography, remote sensing, medical imaging, etc. Quite many techniques of lossy compression have been designed including both standard ones as JPEG [1] and JPEG2000 [2], fractal coders [3] and recently proposed methods as AGU [4], ADCT [5], their modifications and others.

One of the main questions that arises in practice of lossy image compression is how to control output quality of compressed images? Note that classical quantitative criteria (metrics, quality indices) as, e.g., mean square error (MSE) or peak signal-to-noise ratio (PSNR) strictly related to MSE often do not adequately describe compression efficiency especially if visual quality is of prime importance. To characterize visual quality more precisely and adequately, a lot of novel, so called HVS-based, quality metrics has been designed recently including MSSIM[6], WSNR[6], PSNR-HVS [7], PSNR-HVS-M [7] and others. Thus, there are several compression techniques (in particular, those ones oriented on producing better visual quality compared to standard coders), a set of rather efficient quality metrics, and some recommendations concerning setting the values of these metrics for providing appropriate visual quality, for example, invisibility of introduced distortions [8]. However, there is a need in convenient software tool for setting a required quality according to a given metric. Here we describe an automated software designed for this purpose and give some examples of its use for different applications.

## II. Software structure and facilities

A screen-shot of the designed software is presented in Fig. 1. The program interface includes eight functional blocks. The first block (Coders' list) is intended for coder selection from the available set. There are possibilities to choose the following coders: JPEG, SPIHT, JPEG2000, the coders AGU [4] and ADCT [5], as well as the coders able to take into account some peculiarities of human visual system as AGU-M and ADCT-M.

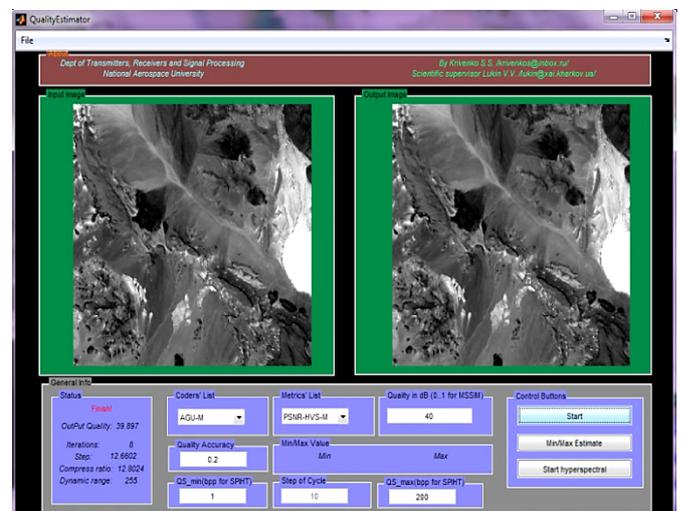

Fig.1 Screen-shot of the software

The second block (Metrics' list) serves for choosing a quality metric according to which output (compressed) image quality control is performed. This block contains sub-blocks for calculating conventional (classical) metric PSNR as well as may visual quality metrics as PSNR-HVS, PSNR-HVS-M, MSSIM, WSNR. In fact, this block can be easily modified (if necessary) by adding any metric available in [6].

The third block (Quality) is intended for setting a desired quality of the output image. Note that for the metrics PSNR, PSNR-HVS, PSNR-HVS-M, WSNR quality is set in dB whilst for the metric MSSIM a value to be set varies from 0 (very bad quality) to 1 (perfect quality).

The fourth block (Quality accuracy) sets accuracy of providing a required quality for a compressed image according to a chosen metric for iterative procedures [9]. It is very difficult (and sometimes even impossible) to provide a required quality absolutely exactly. Thus, accuracy parameter $\Delta$ is to be set based on reasonable considerations. In particular, for the metrics expressed in dB, $\Delta$ not larger than 0.1 dB can be set without any noticeable changes in visual quality.

The fifth block (QS_min/QS_max) sets minimal and maximal quantization steps (scaling factors) for coders where this parameter controls compression characteristics. For SPIHT and JPEG2000, minimal and maximal bpp (bits per pixel) values (from 0 to 8) are set instead of quantization step to vary CR.

The sixth block (Control buttons) serves for program control and contains the following elements: Start – the button for starting compression of iterative procedure, Min/Max Estimate – to estimate a range of a chosen metric variation determined by maximal and minimal values of a parameter controlling compression (quantization step, scaling factor, bpp), Start hyperspectral – to initialize the process of multichannel, in particular, hyperspectral image compression carried out component-wise (sub-band by sub-band), this mode basically serves for applying the software tool to compressing remote sensing data.

The seventh block (Input/Output image) contains input and output images. They are simultaneously presented at monitor screen (as shown in Fig. 1) in order to analyze and compare the results, to study possible compression artifacts, etc.

The eighth block contains quantitative results for a compressed image and compression process. They can be the following: provided quality of the compressed image according to a chose metric, number of iterations this quality has been ensured, finally obtained value of a parameter that controls compression, obtained compression ratio, etc.

Besides, one can optionally use auxiliary functions as format converting (e.g., into RAW-format), direct and inverse homomorphic transformations of data. This can be useful in compressing hyperspectral images that have considerably different dynamic ranges for sub-band data.

## III. Results of software exploitation

The main advantage of the designed software is that it allows compressing an image with a required (according to a chosen metric) level of distortions introduced. Two iterative procedures have been implemented. The first is based on median split applied to initial range of parameter controlling compression. The second one also presumes multiple compression decompression but it starts from a recommended value of control parameter and applies linear interpolation at the final stage [10].

For the second (latter) iterative procedure, the number of iterations is not large (including the case of Hyperspectral data component-wise compression) and it very rarely exceeds 5 whilst is approximately equal to 3 on the average. This allows compressing images with a required quality quite fast. Let us give some examples for remote sensing and digital imaging and medical imaging applications (Figures 2-5). For one sub-band of AVIRIS Hyperspectral data (Fig. 2) compressed by the coder AGU-M with providing PSNR-HVS-M=40 dB, CR equals to 12.64 for scaling factor equal to 12.43. Compression has been carried out in three iterations. For the image Airfield obtained by airborne optical sensor, CR=8.48 has been provided by the same coder with the final scaling factor equal to 11.33, compression has been performed in two iterations (Fig. 3).

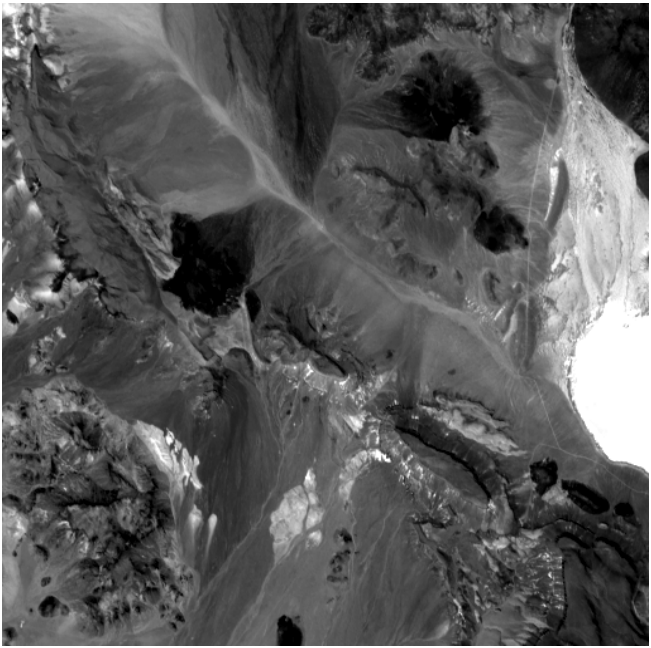

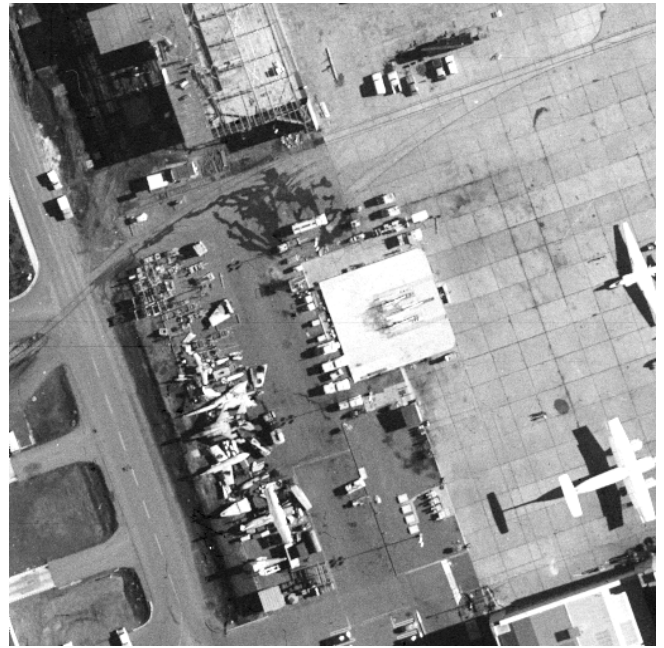

a

a

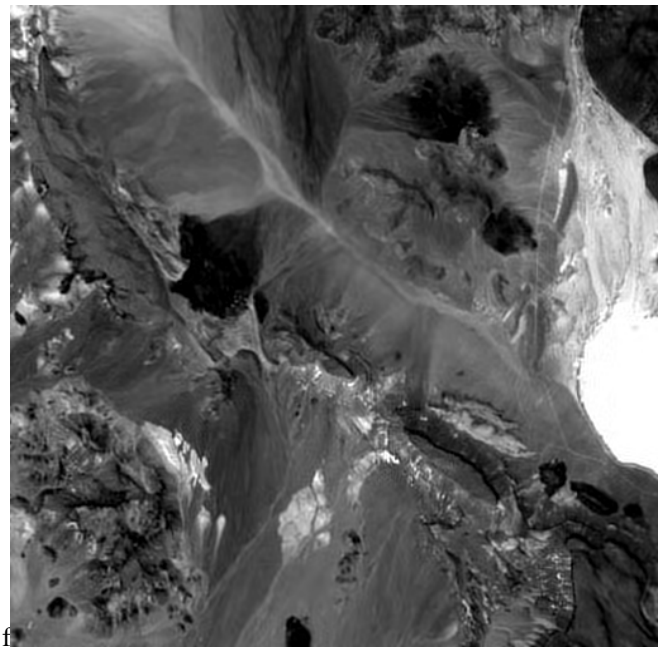

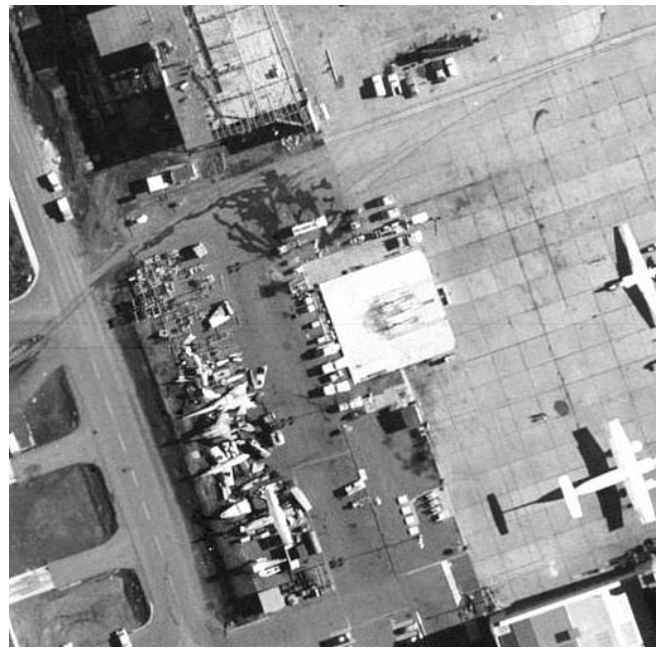

b

b

Fig.2 Sub-band image of hyperspectral remote sensing system AVIRIS Lunar Lake before (a) and after (b) lossy compression by the coder AGU-M, PSNR-HVS-M=40 dB

Fig.3 Grayscale aerial photo Airfield before (a) and after (b) lossy compression by the coder AGU-M with providing a required PSNR-HVS-M=40 dB

Note that PSNR-HVS-M=40 dB as well as MSSIM=0.985 approximately corresponds to the distortion invisibility threshold. For the picture Pole compressed by ADCT-M with providing MSSIM=0.985 (Fig. 4) the attained CR=41.38 for scaling factor 28.9 (4 iterations).

The described approach performs well for medical images as well. For the MRT image presented in Fig. 5 the coder ADCT-M provides PSNR-HVS-M=40 dB with CR=41.41 if the scaling factor is equal to 19.28 where compression is carried out using only three iterations.

The presented results demonstrate that the provided CR can be rather large although introduced distortions are practically invisible (or can be hardly noticed by visual inspection). In this sense, the coders AGU-M and ADCT-M implemented in our software produce CR by 30…60% larger than JPEG and JPEG2000.

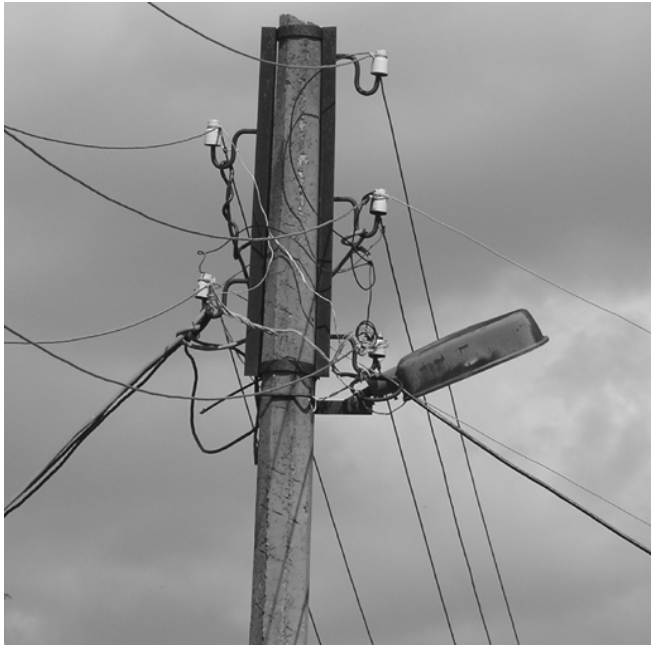

a

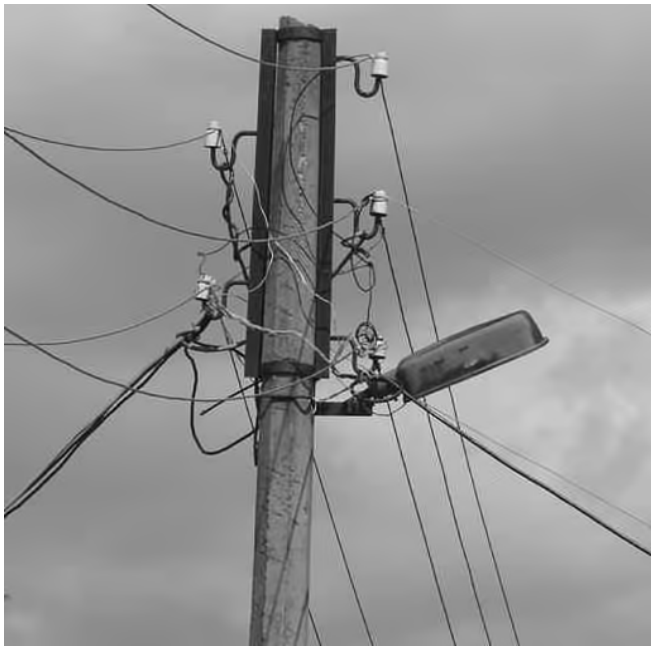

b

Fig.4 Digital grayscale image Pole before (a) and after (b) compression by ADCT-M, MSSIM=0.985

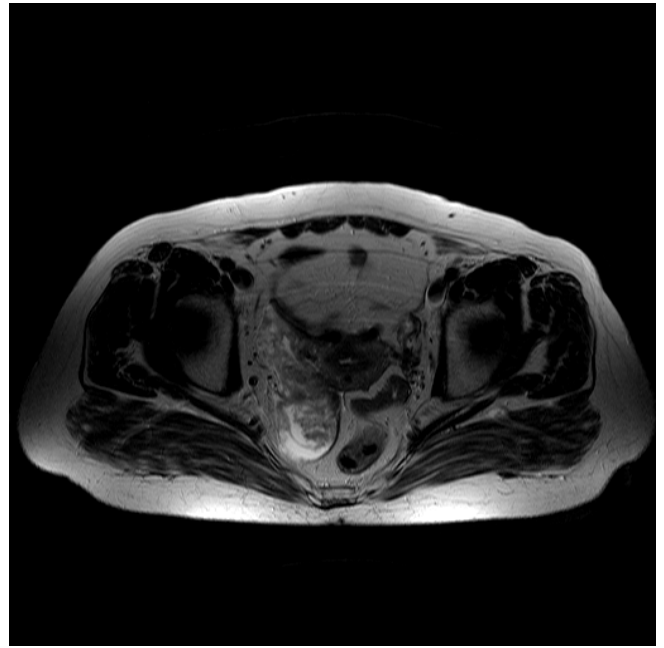

a

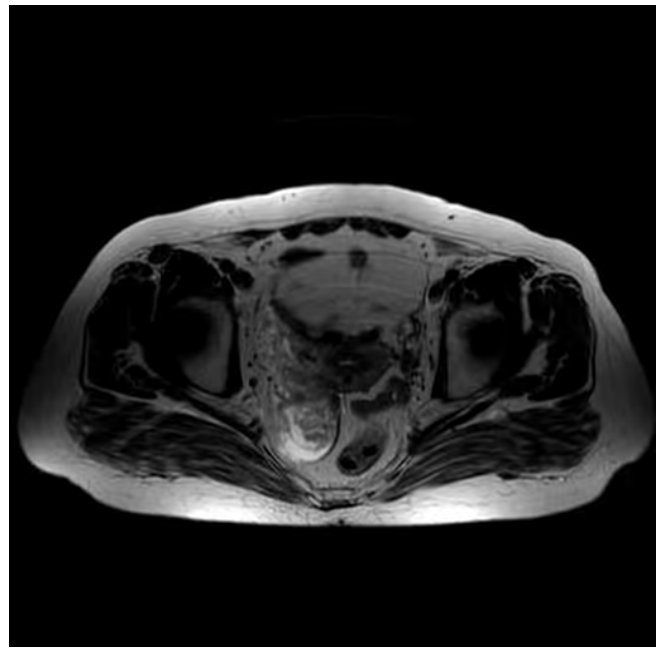

b

Fig.5 Medical MRT image before (a) and after (b) lossy compression by ADCT-M, PSNR-HVS-M=40 dB

IV. Conclusions

The presented data and examples demonstrate the following. The designed software can be successfully used for compressing images of different origin. It allows quite fast (in a few iterations) compression of a given image according to a wide set of quality metrics and their preset values. Several coders can be applied and their performance can be easily compared for a given metric, its preset value (or a set of values) and a given image (or a set of images). This can lead to certain conclusions and practical recommendations. The compressed images can be visualized to control a provided quality, introduced distortions and artifacts (if they are present and visible). The software can be easily advanced (modified) if new coders and/or metrics will appear in the future and their implementations will be freely available.